\documentclass{ws-procs975x65}
\catcode`\@=11 

\global\newcount\chapno \global\chapno=0
\global\newcount\secno \global\secno=0
\global\newcount\meqno \global\meqno=1

\def\newchap#1{\global\advance\chapno by1
\global\secno=0\eqnres@t
\xdef\chapsym{\the\chapno.}
\chapter{#1}
}

\def\newsec#1{\global\advance\secno by1
\global\subsecno=0\eqnres@t
\section{ #1}
}
\def\eqnres@t{\xdef\secsym{\the\secno.}\global\meqno=1}
\def\sequentialequations{\def\eqnres@t{\bigbreak}}\xdef\secsym{}
\global\newcount\subsecno \global\subsecno=0
\def\subsec#1{\global\advance\subsecno by1
\subsection{#1}}

\def\draftmode{\message{ DRAFTMODE }
\writelabels
 {\count255=\time\divide\count255 by 60 \xdef\hourmin{\number\count255}
  \multiply\count255 by-60\advance\count255 by\time
  \xdef\hourmin{\hourmin:\ifnum\count255<10 0\fi\the\count255}}}
\def\nolabels{\def\wrlabeL##1{}\def\eqlabeL##1{}\def\reflabeL##1{}}
\def\writelabels{\def\wrlabeL##1{\leavevmode\vadjust{\rlap{\smash%
{\line{{\escapechar=` \hfill\rlap{\tt\hskip.03in\string##1}}}}}}}%
\def\eqlabeL##1{{\escapechar-1\rlap{\tt\hskip.05in\string##1}}}%
\def\reflabeL##1{\noexpand\llap{\noexpand\sevenrm\string\string\string##1}}}
\nolabels

\def\eqn#1#2{
\xdef #1{(
\secsym\the\meqno)}
\global\advance\meqno by1
$$#2\eqno#1\eqlabeL#1
$$}

\def\eqalign#1{\null\,\vcenter{\openup\jot\m@th
  \ialign{\strut\hfil$\displaystyle{##}$&$\displaystyle{{}##}$\hfil
      \crcr#1\crcr}}\,}

\def\foot#1{\footnote{#1}} 

\global\newcount\refno \global\refno=1
\newwrite\rfile
\def\ref{[\the\refno]\nref}
\def\nref#1{\xdef#1{[\the\refno]}
\ifnum\refno=1\immediate\openout\rfile=refs.tmp\fi
\global\advance\refno by1\chardef\wfile=\rfile\immediate
\write\rfile{\noexpand\bibitem{#1}}\findarg}
\def\findarg#1#{\begingroup\obeylines\newlinechar=`\^^M\pass@rg}
{\obeylines\gdef\pass@rg#1{\writ@line\relax #1^^M\hbox{}^^M}%
\gdef\writ@line#1^^M{\expandafter\toks0\expandafter{\striprel@x #1}%
\edef\next{\the\toks0}\ifx\next\em@rk\let\next=\endgroup\else\ifx\next\empty%
\else\immediate\write\wfile{\the\toks0}\fi\let\next=\writ@line\fi\next\relax}}
\def\striprel@x#1{} \def\em@rk{\hbox{}} 
\def\lref{\begingroup\obeylines\lr@f}
\def\lr@f#1#2{\gdef#1{\ref#1{#2}}\endgroup\unskip}

\def\addref#1{\immediate\write\rfile{\noexpand\item{}#1}} 
\def\startrefs#1{\immediate\openout\rfile=refs.tmp\refno=#1}
\def\xref{\expandafter\xr@f}\def\xr@f[#1]{#1}
\def\refs#1{\count255=1[\r@fs #1{\hbox{}}]}
\def\r@fs#1{\ifx\und@fined#1\message{reflabel \string#1 is undefined.}%
\nref#1{need to supply reference \string#1.}\fi%
\vphantom{\hphantom{#1}}\edef\next{#1}\ifx\next\em@rk\def\next{}%
\else\ifx\next#1\ifodd\count255\relax\xref#1\count255=0\fi%
\else#1\count255=1\fi\let\next=\r@fs\fi\next}
\newwrite\lfile
{\escapechar-1\xdef\leftbracket{\string\{}
\xdef\rightbracket{\string\}}}

\def\writestop{\def\writestoppt{\immediate\write\lfile{\string\pageno%
\the\pageno\string\startrefs\leftbracket\the\refno\rightbracket%
\string\def\string\secsym\leftbracket\secsym\rightbracket%
\string\secno\the\secno\string\meqno\the\meqno}\immediate\closeout\lfile}}

\catcode`\@=12 
\def\inv{^{\raise.15ex\hbox{${\scriptscriptstyle -}$}\kern-.05em 1}}

\def\Dsl{\,\raise.15ex\hbox{/}\mkern-13.5mu D} 
\def\dsl{\raise.15ex\hbox{/}\kern-.57em\partial}

\def\lspace{\ifx\answ\bigans{}\else\qquad\fi}
\def\lbspace{\ifx\answ\bigans{}\else\hskip-.2in\fi} 
\def\boxeqn#1{\vcenter{\vbox{\hrule\hbox{\vrule\kern3pt\vbox{\kern3pt
	\hbox{${\displaystyle #1}$}\kern3pt}\kern3pt\vrule}\hrule}}}
\def\tilde{\widetilde}  
\def\ph{{\phi}}

\def\tt{\tilde{\theta}}

\def\back{{{\raise.4em\hbox{$\, _\backslash\,$}}}}

\font\black=msbm10 scaled\magstep1
 2

\def\field #1{\hbox{{\black #1}}}

\def\frac#1#2{{#1\over #2}}

\def\big R{{\hbox{{\bigfield R}}}}
\def\bbig R{{\hbox{{\bbigfield R}}}}

\font\af=msbm10

\font\bbm=bbm10
\def\Z{\hbox{\af Z}}

\def\SS{\hbox{$\bbm \Delta$}}
\mathchardef\imath="717B
\def\sc{\af}
\def\inbar{\,\vrule height1.5ex width.4pt depth0pt}
\def\IB{\relax{\rm I\kern-.18em B}}
\def\IC{\relax\hbox{$\inbar\kern-.3em{\rm C}$}}
\def\ID{\relax{\rm I\kern-.18em D}}
\def\IE{\relax{\rm I\kern-.18em E}}
\def\IF{\relax{\rm I\kern-.18em F}}
\def\IG{\relax\hbox{$\inbar\kern-.3em{\rm G}$}}
\def\IH{\relax{\rm I\kern-.18em H}}
\def\II{\relax{\rm I\kern-.18em I}}
\def\IK{\relax{\rm I\kern-.18em K}}
\def\IL{\relax{\rm I\kern-.18em L}}
\def\IM{\relax{\rm I\kern-.18em M}}
\def\IN{\relax{\rm I\kern-.18em N}}
\def\IO{\relax\hbox{$\inbar\kern-.3em{\rm O}$}}
\def\IP{\relax{\rm I\kern-.18em P}}
\def\IQ{\relax\hbox{$\inbar\kern-.3em{\rm Q}$}}
\def\IR{\relax{\rm I\kern-.18em R}}
\font\cmss=cmss10 \font\cmsss=cmss10 at 10truept
\def\IZ{\relax\ifmmode\mathchoice
{\hbox{\cmss Z\kern-.4em Z}}{\hbox{\cmss Z\kern-.4em Z}}
{\lower.9pt\hbox{\cmsss Z\kern-.36em Z}}
{\lower1.2pt\hbox{\cmsss Z\kern-.36em Z}}\else{\cmss Z\kern-.4em Z}\fi}
\def\IGa{\relax\hbox{${\rm I}\kern-.18em\Gamma$}}
\def\IPi{\relax\hbox{${\rm I}\kern-.18em\Pi$}}
\def\ITh{\relax\hbox{$\inbar\kern-.3em\Theta$}}
\def\IOm{\relax\hbox{$\inbar\kern-3.00pt\Omega$}}

\def\ha{{1\over2}}
\def\Asl{A \kern-1.9mm {/} \kern.4mm}
\def\parsl{\raise.15ex\hbox{/}\kern-.57em\partial}
\def\pasl{\partial \kern-1.45mm {/}}
\def\Dsl{D \kern-2.33mm \raise.18ex\hbox {/} \kern.4mm}
\def\Dsla{D \kern-2.43mm \raise.20ex \hbox{/} \kern.4mm_{\af A}}
\def\Fsl{F \kern-2.33mm {/} \kern.4mm}
\def\rsl{r \kern-2.33mm {/} \kern.4mm}
\def\psl{p \kern-2.33mm {/} \kern.4mm}

\font\af=msbm12

\def\cs{{Chern-Simons}}
\def\cst{{Chern-Simons theory}}
\def\be{\begin{eqnarray}}    
\def\ee{\end{eqnarray}}
\begin{document}
\input epsf.tex
\nref\Deser{J. Schonfeld,  Nucl. Phys.  {\bf B 185} (1981) 157
 S. Deser, R. Jackiw, S. Templeton, Phys. Rev. Lett. { \bf 48}
(1982)    975;   Ann.  Phys. (N.Y.) {\bf  140} (1982) 372}

\nref\runo
{R. Jackiw,
In { \it Gauge Theories of the Eighties}, Eds. E.
Ratio, J. Lindfords, Lecture Notes in Physics, vol. 181, Springer
(1983) }

\nref\ram{
M. Asorey, P.K. Mitter, Phys. Lett. {\bf B153} (1985) 147}

\nref\pisrao{R. D. Pisarski, S. Rao,
 Phys. Rev. {\bf D 32} (1985) 2081}\nref\sem{W.
Chen,  G. W. Semenoff, Y.-S. Wu, Mod. Phys. Lett. {\bf A5}  (1990)
1833}
\nref\LAG{L. Alvarez-Gaum\'e, J.M.F. Labastida and A.V.
Ramallo, Nucl. Phys. {\bf B334} (1990) 103}
\nref\rafa{ M. Asorey, F. Falceto,
Phys. Lett. {\bf B 241} (1990) 31}
\nref\carmelo{
C.P.  Martin, Phys.  Lett. {\bf B 241} (1990) 513}
\nref\shifman
{M. A. Shifman, Nucl. Phys. {\bf B352} (1991) 87}

\nref\usb{M. Asorey, F. Falceto, J.L. 
L\'opez and G. Luz\'on, Nucl. Phys. {\bf B429} (1994) 344.}

\nref\aps{M. Atiyah, V. Patodi, I.M. Singer, 
I. Math. Proc. Comb. Phil. Soc. {\bf 77}
(1975) 43 and II {\bf 78} (1975) 405}

\nref\redlich{A.V. Redlich, Phys. Rev. Lett. {\bf 52} (1981) 18; Phys. Rev. 
{\bf D29} (1984) 2366.}

\nref\dunne{ G. Dunne, K. Lee and Ch. Lu, Phys. Rev. Lett. {\bf 78} (1997)
3434}

\nref\fosco{C. Fosco, G.L. Rossini and F. A. Schaposnik,
Phys. Rev. Lett. {\bf 79} (1997) 1980}

\nref\deser{S. Deser, L. Griguolo and D. Seminara,
Phys. Rev. Lett. {\bf 79}
(1997) 1976;  Phys. Rev. {\bf D 57}
(1998) 7444}

\nref\jack{R. Jackiw and Y.S. Pi,  Phys.  Lett. {\bf B 423}  (1998)  364.}

\nref\witten{E. Witten, Commun. Math. Phys. {\bf 121}  (1989) 351.}

\nref\nat{D.
Bar--Natan, E. Witten, Commun. Math. Phys. {\bf  141} (1991) 423}

\nref\GK{
K. Gawedzki and A. Kupiainen, Phys. Lett. {\bf B215} (1988) 119;
Nucl. Phys. {\bf B320} (1989) 649}

\nref\slavnov{A. Slavnov, Theor. Math. Phys. {\bf 33} (1977) 210}




\nref\aflbis{M.Asorey and F. Falceto, unpublished (1990)}

\nref\korchemski{G. P. Korchemsky, Mod. Phys. Lett. {\bf A6} (1991) 727}

\nref\kimura{T. Kimura, Prog. Theor. Phys. {\bf 92} (1994) 693.}

\nref\narayanan{R. Narayanan and J. Nishimura, Nucl. Phys. {\bf B 508}
(1997)  371.}

\nref\slavnew{A. Slavnov,  Phys. Lett. {\bf B415} (1997) 390}


\nref\wwitten{E. Witten,Phys. Lett. {\bf B 117} (1982) 324.}

\nref\nair{D. Karabali and V.P. Nair, Nucl. Phys. {\bf B464}  (1996) 135; 
Phys. Lett. {\bf B379} (1996) 141; Int. J.  Mod. Phys {\bf A12} (1997) 1161.}

\nref\teper{
M. Teper, { Phys. Lett.} {\bf B311} (1993)  223; O. Philipsen,
M. Teper and H. Wittig, Nucl. Phys. {\bf B469}  (1996) 445 ;
M. Teper, Phys. Rev. {\bf D59} (1999) 014512}

\nref\atiyah{M. F. Atiyah and I.M. Singer, Ann. of Math. {\bf 87} (1968) 485, 
546; {\bf 93} (1971) 1, 119, 139; M. F. Atiyah and G.B. Segal, 
Ann. of Math. {\bf 87} (1968) 531.}

\nref\pietra{L. Alvarez-Gaum\'e, S. Della Pietra and G. Moore, Ann. Phys.
{\bf 163} (1985) 288.}

\nref\seminara{D. Seminara, In {\it Particles, Strings and Cosmology},
Ed. P. Nath, World Sci., Singapore (1999)}

\nref\jllp{
J.L. L\'opez, J. Phys. {\bf A  31}  (1998) 7955}

\nref\kar{M. Asorey, J. Geom. Phys.  {\bf 11}(1993) 63}

\nref\afll{M. Asorey, F. Falceto,  and G. Luz\'on,
Phys. Lett. {\bf B 349} (1995) 125 }

\nref\affs{M. Asorey and  F. Falceto,
Phys. Rev. Lett. {\bf 77}(1996) 3074 }

\nref\bosnair{M. Bos, V.P. Nair, Phys. Lett. {\bf B 223} (1989) 61}

\nref\wittten{E. Witten, in {\it Physics and Mathematics of
Strings},   L. Brink ed., World Sci., Singapore (1990).}








\nref\AS{S.  Axelrod , I.M. Singer,
In {\it Proceedings Differential geometric methods in theoretical physics}, 
vol. {\bf 1},  S. Catto and A. Rocha, eds., World Sci, Singapore (1992) 3;
 Diff. Geom. {\bf 39} (1994) 173}

\nref\achucarro{A. Ach\'ucarro and P. Townsend,Phys. Lett. 
{\bf B 180} (1986)383}

\nref\flc{A. Capelli, D. Friedan and J. I. Latorre, Nucl. Phys. {\bf 352} (1991) 616 }



\overfullrule=0pt

\title{\Large  Non-analyticities in \vskip8pt three-dimensional gauge theories}
\author{ M. Asorey\foot{\uppercase{I} am one of the very fortunate persons 
who had deep
scientific and vital resonances with \uppercase{I}an \uppercase{K}ogan. 
\uppercase{S}till under the effect
of the tragedy, in \uppercase{J}une 2003, \uppercase{I} promised  to \uppercase{I}an and to myself to finish a 
joint paper which we had outlined few weeks before. \uppercase{T}here is an old spanish
popular aphorism telling that {\it  Lo que no puede ser, no puede ser, y adem\'as 
es imposible} (what cannot be, cannot be, and furthermore it is impossible!).
\uppercase{N}ow \uppercase{I}  better understand the meaning of the 
tautological aphorism, 
without \uppercase{I}an's resonances \uppercase{I} will not be able to accomplish my promise.
\uppercase{I} am sorry, \uppercase{I}an.}$\,$ ,  D. Garcia-Alvarez}

\address{\it Departamento de F\'{\i}sica Te\'orica,
Universidad de Zaragoza,  50009 Zaragoza, Spain.}


\author{ J.L. L\'opez}

\address { Departamento de Matem\'aticas e Inform\'atica, Universidad P\'ublica
de Navarra, 
31006 Pamplona, Spain.}

\maketitle

\abstracts{
Quantum fluctuations generate in three-dimensional gauge theories
not only radiative corrections  to the Chern-Simons  coupling 
  but also non-analytic terms in the effective action.
We review  the role of those terms  in  gauge theories with  massless
fermions and Chern-Simons theories. The explicit form of 
non-analytic terms turns out to  be
dependent on the regularization scheme and in consequence
the very existence of phenomena like parity and framing anomalies becomes
regularization dependent. In particular we find  regularization
regimes where both anomalies are absent. Due to the presence of  non-analytic terms
the effective action becomes not only discontinuous but also singular for
some background gauge fields which include sphalerons.
The appearence of this type of singularities is linked to the
existence of nodal configurations in physical states and 
tunneling suppression at some classical field configurations.
In the topological field theory the number of physical states may also
become regularization  dependent.  Another consequence of the peculiar
behaviour of three-dimensional theories under parity odd regularizations
is the existence of  a simple mechanism of generation of a mass gap in pure Yang-Mills
theory by a suitable choice of regularization scheme. The generic value
of this mass does agree with the values obtained in  Hamiltonian and
numerical analysis. Finally, the existence of different regularization
regimes unveils the difficulties of establishing a Zamolodchikov
c-theorem for three-dimensional field theories in terms of the induced gravitational
Chern-Simons couplings.}

\overfullrule=0pt  \hyphenation{systems}
\bigskip\vfill \noindent\baselineskip=16pt plus 2pt minus 1pt

\date{ }  
\newsec{Introduction}
In 2+1 dimensions the number of degrees of freedom of massive and massless
relativistic  particles is the same. This peculiar behaviour permits a smooth
transition from  massless  to   massive regimes in the same theory without the need of
extra fields. In gauge theories this transition can be simply achieved by the
addition of  a 
Chern-Simons term to the ordinary Yang-Mills action \Deser. For the same
reasons there is no protection against the existence of radiative quantum corrections  
which either generate or suppress  the topological mass.

The special characteristics  of Chern-Simons term and its peculiar  transformation law
 under  large gauge transformations requires the quantization    of its coupling
constant $k\in \IZ$  when the gauge group is compact.  The
constraint arises in the
covariant formalism
as a consistency condition for the definition of the
euclidean functional integral
due to the special transformation properties
of the Chern-Simons action under
large gauge transformations \Deser.
In the canonical formalism it appears as a necessary condition for the
integration of Gauss law  on the physical states \runo\ram.
Both interpretations are based on non-infinitesimal symmetries and therefore the quantization
condition can not be inferred from perturbative arguments. However,
unexpectedly the perturbative contributions of quantum
fluctuations do not seem to change the integer nature of the  Chern-Simons coupling
constant in most of   standard renormalization schemes
\pisrao
--\usb. 
From a pure quantum field theory point of view this
behavior is bizarre because in absence of perturbative symmetry
constraints   there must always exist regularization schemes where the effective values
of the coupling constants of marginal local terms are arbitrary. Indeed,
such   regularization schemes exist   but require a fine tuning
of the leading ultraviolet behavior  of parity even  and parity odd
terms of regulators \usb .

 The perturbative quantum corrections are not the only contributions
of quantum fluctuations.  There exist additional
contributions to the effective gauge action
which cannot be obtained in perturbation theory because
they  are not analytical on gauge fields. The presence of such
non-analytic contributions  in one loop approximation
is more evident in the case of regularizations which do not preserve the
integer value character of the effective Chern-Simons coupling.
They appear as necessary  to compensate the anomalous transformation law of  Chern-Simons
terms under large gauge transformations.  The role of those terms is crucial to understand
the finite temperature behaviour of gauge theories in $2+1$ dimensions
\dunne--\jack. 
They are similar   to the well known non-analytic terms
which appear in the  $\eta$--invariant of the spectral asymmetry \aps\  of the operator
 $\ast d^{\ }_A+d_A\ast$  induced by the changes of signs in the spectral flow \redlich.

The study of  non-analytic terms of the effective action
and their physical implications is the
main goal of our analysis. The discontinuities associated to
these terms yield singularities which in the case
of  Chern-Simons theory seem to be mere artifacts of perturbation theory. The
origin of the singularities is the same that  appears
 in ordinary gauge theories in presence of massless quarks  in
the fundamental representations. In this case  the singularities do have a
simple physical origin, the existence of zero-modes of Dirac operator.

The main result of the paper is the proof  that this kind of non-analyticities are
regularization dependent which  provides a further support to the claim that
 different renormalization  methods define in fact different physical
theories. The perturbative corrections to the
\cs\  coupling constant  can  also be different and depend on the
regularization  method  but those
differences can be compensated  in general by the choice of  different  
renormalization schemes. However, the presence of 
different non-analytic contributions cannot be
changed by the choice of appropriate  renormalization schemes. In some way
this provides a physical meaning  to  the non-perturbative constraint  that
requires  the coupling of  \cs\ counterterms must be an integer  value. 
The  meaning of the restriction  is  that   the analytic
behaviour of the effective partition function cannot be changed by the
renormalization scheme and provides a novel physical role to the choice 
of regularization method.  

Parity anomaly and framing anomaly have  a common origin in the existence
of odd quantum effects. Because of their dependence
on the regularization method it is possible, thus, to find out  some
regularization regimes where both anomalies are absent.

Finally, the regularization dependence of these phenomena is also responsible of the failure of  simple
attempts to define  a Zamolodchikov's c-function in terms of gravitational \cs\
terms in order to generalize of Zamolodchikov c-theorem for  three-dimensional 
theories.


\newsec{Chern-Simons theory}
In the limit of infinite topological mass  the gauge theory reduces to a
Chern-Simons topological theory \witten\  governed by the action
$$              
{ k}\  S_{\rm CS}(A) = {{ k}\over 4\pi}
\int_{M}{\rm Tr}\left(A\land
 dA +{2\over 3}A\land A\land A\right) \qquad    k\in 
{\field Z}
$$
where  the coupling constant $k$  must be an integer for
compact groups to have a consistent quantization \foot{
A generalization for non-compact gauge groups is straightforward
\nat.}. Let us consider $SU(N)$ gauge field theories for simplicity.

The theory is superenormalizable also in this topological limit. In the 
Hamiltonian formalism divergences appear in the normalization of physical 
states and the hermitian product of the Hilbert space \GK. The removal of
these divergences generates a shift in the renormalized Chern-Simons coupling
constant $k_{\rm R}= k+   N$. In the covariant
formalism the propagator is very singular because of the large gauge 
symmetry of the theory originated by its topological character. 
In perturbation theory one way of improving the UV behaviour of
the propagator without breaking gauge invariance is by
introducing higher derivative regulating terms into the classical action, e.g.
$$
S_{\rm \Lambda}(A)= S_{\rm CS}(A)+ S^+_{\rm R}(A)$$
$$
S_{\rm R}^+(A)={\lambda_+\over \Lambda}
\int_{M}{\rm Tr}\, F_{\mu\nu}(A) \left({\mathbb{ I}}+ 
{\bbm{\Delta}_A \over \Lambda^2}\right)^m F_{\mu\nu}(A),
$$
where  $\bbm{\Delta}_A=d^\ast_Ad^{\phantom \ast}_A+d^{\phantom \ast}_Ad^\ast_A $ is the covariant laplacian.
For large enough values of the exponent 
$m$ there are not UV superficial divergences in diagrams with more than
one loop. However, one loop divergences need of an extra  Pauli-Villars
regularization \slavnov. The resulting one loop effective action  has no
divergences even after the removal of the ultraviolet regulator $\Lambda\to\infty$.
The renormalized perturbative effective action is of the form
\eqn\scalar{{\rm \Gamma}^{\rm pert}(A^R)={\rm \Gamma}_R(A^R)+ 
i {\rm \Gamma}_I(A^R)$$
$${ \Gamma_{I}(A^R)}={\black k_{\rm R}}
\, S_{\rm CS}(A^R)}
with $k_{\rm R}= k+  N$. The first non-trivial contribution to  ${\rm \Gamma}(A^R)$
arises from the four point function \aflbis.

The  Hamiltonian approach yields similar results, but
this coincidence is not based on general symmetry principles. Thus, it should be possible
the existence of a regularization where the renormalization of $k$ is not
a simple shift  of $k$ by N units. Indeed, there exist other gauge invariant 
regularizations, e.g. \usb.
$$
S_{\rm \Lambda}(A)= S_{\rm CS}(A)+ S^-_{\rm R}(A)$$
$$
S_{\rm R}^-(A)={\lambda_-\over \Lambda^2}
\int_{M}{\rm Tr}\, {   \epsilon^{\alpha\sigma\mu}}\, 
F_{\alpha\nu}(A) \left(\mathbb{I}+ 
{\bbm{\Delta}_A \over \Lambda^2}\right)^n
D^\sigma_A\left(\mathbb{I}+ 
{\bbm{\Delta}_A \over \Lambda^2}\right)^n \! F_{\mu\nu}(A)
$$
which  after removing  one-loop divergences yield an effective
action like \scalar, but without  radiative contributions  to the
 effective value of the coupling constant $k_{\rm R}= k$.
 Even more general
regularizations can be conceived, e.g.
$$S_{\rm \Lambda}(A)= S_{\rm CS}(A)+ S^+_{\rm R}(A)+
 S^-_{\rm R}(A)$$
In that case the result depends on the relative weights  $\lambda_->0$ and
$\lambda_+>0$\foot{If $\lambda_-<0$ the results are slightly different \usb } 
of  $S^+_{\rm R}$
and $S^-_{\rm R}$
$$k_{\rm R}= \begin{cases}
{ 
{k +   N}\phantom{\Big[}}& {{\rm if}\  m > 2n+1/2}\cr
 {k +    {2 N \over\pi}
\arctan{\lambda_+\over\lambda_- }}& {{\rm if}\     m = 2n+1/2}\cr
{{k} \phantom{\Big[}}& {{\rm if}\   m < 2n+1/2 }\cr
\end{cases}
$$

In these very general regularization schemes the radiative corrections to
the coupling constant present three different regimes which depend on the
interplay between the ultraviolet behaviors of parity even terms $S^+_{\rm R}$ of the regularized
action and the parity odd terms of $S^-_{\rm R}$.

  In the first regime
the leading ultraviolet terms are parity even. The
effective \cs\
coupling constant gets  shifted by $N$ ( $k\to k+N$), due to
one loop gluonic radiative corrections. The third regime is
characterized by an ultraviolet behavior dominated by parity odd terms
 and the absence of radiative corrections to $k$. 
In the transition regime parity even and parity odd
terms have the same ultraviolet behavior and  the
quantum corrections to $k$ can take any real value
which depends  on the relative coefficients of the
leading terms of parity even and parity odd interactions.

The phenomenon
can be pictorially  understood by looking at the way the shift of $k_R$ is
generated. In fact, 
$${   k_R}=k+ {2N\over\pi}
\int_{0}^{\infty}{d\Phi \over 1+\Phi^2}=k+{N\over \pi}\arctan \Phi(\infty) $$
and the behaviour of 
$$\Phi={\lambda_+ p(1+p^2)^m\over
1+\lambda_- p^2(1+p^2)^{2n}}$$
 is dictated by the form of  $S_{\rm \Lambda}(A)$
\medskip

\begin{figure}[ht]
\centerline{\epsfxsize=10cm \epsfbox{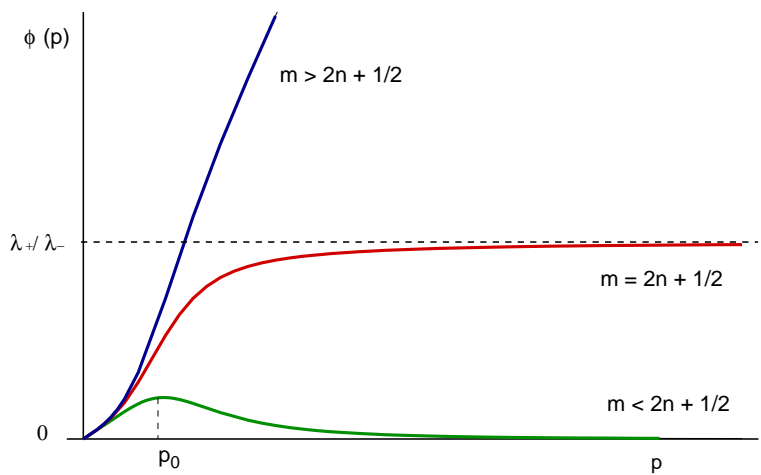}}
\caption{ Behaviour of the  function $\Phi$ for different
regularization  regimes}
\end{figure}

The actual value  of the effective
coupling
constant can always be modified by  a different choice of
renormalization
scheme because the Chern-Simons term is local and can be added as a conterterm. 
However, as pointed out in the previous section, the behaviour
of Chern-Simons term under large gauge transformations
requires that the bare coupling constant $k$ must be an integer number
otherwise the quantum theory will be inconsistent, e.g.  
the functional integral  will be ill defined.
Such a constraint is a pure  non-perturbative requirement, 
because large gauge transformations
map small fields into large gauge fields and, therefore,
they are genuine non-perturbative symmetries.
In consequence, although in perturbation
theory any  local BRST
invariant counterterm is valid, only  counterterms which
preserve the non-perturbative consistency condition
can be added to the bare action. The condition
imposes a very stringent
constraint on  counterterms which have to preserve the
integer valued character of the bare coupling constant $k$.
In particular, if the
effective value of $k_R$ is not an integer, one cannot
 reduce the
physical behavior of the system to the standard integer
valued case
by a consistent renormalization.
Thus, the first and third regularization schemes are generic and equivalent
from the physical point of view but
 the transition regime $m=2n+\ha$ can not be
reduced any of the
other two regimes by the choice of  a different scheme renormalization.
In fact, the regime defines  a new different theory.

In the generic case there is a correspondence of Chern-Simons states and
the primary fields of  rational conformal field theories \witten. In the
transition regime the corresponding two-dimensional theory
will be non-rational.
In this sense, the transition regularization really defines a new  type of theory,

\newsec {Parity Anomaly}
\medskip
The existence of different regimes in the regularization of Chern-Simons theory
opens new possibilities for the analysis parity anomaly.

This insight is further supported by the  existence of  a straightforward
connection between  one loop corrections of Chern-Simons theory and the determinant
of a massless fermion in the adjoint representation \korchemski. Indeed, the
second variation of Chern-Simons term and the corresponding ghost terms
in a covariant Landau gauge yields an operator $\SS_A$ which is equivalent to
the square of Dirac operator $(\Dsla^{^{ \rm ad}})^2$ for adjoint fermions.
\eqn\ddel{
\Delta_A=\begin{pmatrix}
 {\ast d_A} & {d_A}\\
{d_A^\ast} &{ 0}  
\end{pmatrix}\approx \, 
   (\Dsl_{\af A}^{^{ \rm ad}})^2 
}
Therefore,
$${\det}{ \Dsl_{\af A}^{^{ \rm ad}} }=
 {\rm e}^{-{1\over 2}\Gamma^{[1]}{(A)}}.$$
The effect of the existence of different regularization regimes is more
intriguing because gauge invariance seems to be broken in the transition regime. 
Indeed,   three different regimes  can be generated
 by the following  regularization of Dirac operator
$$
\Dsl_{\sc A}^\Lambda=
{ \Dsl_{\sc A}}+ \lambda_+
{\Dsl_{\sc A}^2 \over \Lambda}
\left(\mathbb{I}+{\Dsl_{\sc A}^2\over 
\Lambda^2}\right)^{m}
+\lambda_-{\Dsl_{\sc A}^3 
\over \Lambda^2}\left(\mathbb{I}+{\Dsl_{\sc A}^2
\over \Lambda^2}\right)^{2n} 
$$
with $\lambda_\pm>0$ and the corresponding Pauli-Villars regulators. In that case the  effective Chern-Simons coupling
behaves in a similar way to the case of pure Chern-Simons theory.

$$k_{\rm R}= \begin{cases}
{{N}{\phantom{\Bigr[ \Bigl]}}} & {{\rm if}\  m > 2n+{1\over 2 }}\cr
{    {2 N \over \pi} \arctan{\lambda_+\over\lambda_- }}& {{\rm if}\     m  = 2n+{1\over 2 }}\cr
{{0} {\phantom{\Bigr[ }}} &{{\rm if}\   m <
2n+ {1\over 2 }}\end{cases}
$$
If the fermions are in the fundamental representation of $SU(N)$ the result
is analogous
$$k_{\rm R}= \begin{cases} 
{{1\over 2}
\phantom{\Bigr[ \Bigl]}}
&  {{\rm if}\  m > 2n+{1\over 2 }}\cr
{    {1 \over\pi} \arctan{\lambda_+\over\lambda_- }} &{{\rm if}\     m  = 2n+{1\over 2 }}\cr
{{0}
{\phantom{\Bigr[ }}}
 &{{\rm if}\   m <
2n+ {1\over 2 }}\end{cases}
$$
Although the Pauli-Villars regularization method used here is completely gauge invariant
also under large gauge transformation,  gauge invariance under those transformations 
seems to be broken in the first two cases  because the effective Chern-Simons term
is not invariant. The puzzle is solved by noticing
that the analytic perturbative radiative corrections do not exhaust all 
quantum corrections to the effective action. In fact, gauge invariance
requires that the full radiative corrections must have a non-analytic conterpart
which permit to recover full gauge invariance. Indeed,
$${\rm \Gamma}^{\rm }(A^R)={\rm \Gamma}_R(A^R)+ 
i {\rm \Gamma}_I(A^R)\qquad{\rm with }\quad{\Gamma^{}_{I}(A)}={ k_{\rm R}}
\, S_{\rm CS}(A)+   h(A)$$
where $ h(A)$  has a non-analytic dependence on $A$. However, in that case 
parity symmetry is not preserved
at the quantum level because $ S_{\rm CS}(A)$ is not invariant under parity
symmetry whereas as it  will be shown  later  $h(A)$ is parity invariant. This fact
is on the  origin of parity anomaly of three-dimensional massless fermions.

However, what is really intriguing  is  that in the third regime $m<2n + 1/2$ there is
no parity anomaly because $k_R=0$ and the theory is at the same time invariant under global
gauge transformations. This means that in fact, contrary to the common wisdom,
the parity anomaly is not an unavoidable physical phenomena
in a gauge invariant framework. 
A similar result is obtained with standard regularizations and infinite number
of  Pauli-Villars fields or lattice regularizations  \kimura\narayanan.
The results are reminiscent of those obtained by Slavnov \slavnew\ for the cancellation of 
the $SU(2)$ global Witten's anomaly in four-dimensional theories with 
chiral fermions in the fundamental representation  
\wwitten.
The main difference between both results is that  the Slavnov method 
requires an infinite number of Pauli-Villars regulating fields to cancel
 the anomaly in a  gauge invariant way, whereas in this
 case a very simple UV modification of fermionic interactions 
yields a similar effect with a finite number of Pauli-Villars fields. 


In the transition regime  parity is also broken but the coefficient of the terms responsible
for this phenomenon are different from those of the case where $m>2n + 1/2$.
The ambiguity in the appearance or not of parity anomaly suggests that
the effect  looks  more like a spontaneous symmetry breaking than
a genuine anomaly breaking.  Perhaps the phenomenon is   nothing 
but a simple example of a more  general feature on the breaking mechanism  
of  discrete symmetries  in three dimensions.  

\newsec{Mass gap in Yang-Mills theory}
Indeed, the same phenomenon arises in 
 the analysis of pure Yang-Mills theory
\eqn\ym{S(A)={1\over 2 g^2}\int_{M}{\rm Tr} |F(A)|^2.}
 Using a similar regularization method which includes parity odd regulating terms and
Pauli-Villars fields  a Chern-Simons term can be induced in pure Yang-Mills theories.
 The infrared behaviour is dominated by the Yang-Mills term \ym\  which is
parity even. The leading UV terms might be either a parity even term 
of the type $S^+_R(A)$ or a parity odd term like  $S^-_R(A)$. 
The general result is 
$$k_{\rm R}= \begin{cases}
{{0}{\phantom{\Bigr[ \Bigl]}}} & {{\rm if}\  m > 2n+{1\over 2 }}\cr
{   - {2 N \over \pi} \arctan{\lambda_+\over \lambda_- }}& {{\rm if}\     m  = 2n+{1\over 2 }}\cr
{{-N} {\phantom{\Bigr[ }}} &{{\rm if}\   m <
2n+ {1\over 2 }}.\end{cases}
$$
In the first case no Chern-Simons coupling is generated whereas in the third case  there is 
a non-trivial Chern-Simons radiative contribution with a coefficient  $k_R=-N$.
Both  results follow from the  behaviour  of the flow displayed in Fig. 1.
In the third case the non-trivial \cs\ term generates a topological mass  $$m={g^2 N \over 2\pi} $$ 
 which  is in agreement with the actual value of the mass
gap in pure Yang-Mills theories \nair\teper.
The generation of a \cs\ term in the pure Yang-Mills theory points out the  instability
of  the renormalization group flow. Moreover, it points toward   a possible
mechanism of generation of a mass gap in pure Yang-Mills theory. In this regime
 the theory is massive but parity symmetry is broken unlike the standard regime
of Yang-Mills theory.

In the transition regime  the theory gets a mass which depends  
 on the relative weights of  the leading parity even and parity odd terms.

%


\newsec{Non-analytic contributions}
The existence of non-analytic contributions to the imaginary part of the
effective action
${\Gamma^{[1]}_{I}(A)}={ k_{\rm R}}
\, S_{\rm CS}(A)+   h(A)$
of massless fermionic determinants is known since the discovery of 
the spectral asymmetry and index theorem \aps\atiyah. In the present case they are pointed
out by the existence of Chern-Simons terms with non-integer 
coefficients \pietra\seminara. The  Pauli-Villars regularization
method  preserves gauge invariance and the only way to ensure the gauge invariance of
the final result is by admitting the existence of a non-analytic contribution in
$h(A)$ which transforms as 
\eqn\h{
h\lbrack A^g\rbrack = h\lbrack A\rbrack +2\pi k_R n
}
under large gauge transformations.

The fermionic determinant $ \Dsla$ is expected to have an analytic dependence on $A$ but the effective
action is the logarithm of this determinant. The existence of a zero in the
determinant induces a singularity in the effective action.  
Thus, the effective action $\Gamma_A$ diverges
for  ({\it nodal}) gauge configurations with fermionic zero modes.
 Every zero of an analytic function has
an integer degree, which is measured by the discontinuity of the 
imaginary part of
the corresponding logarithm. 
 Thus, if the fermionic determinant  has an analytic
dependence on the background gauge field, 
the only possible discontinuities at a 
nodal configurations must by integer multiples of $\pi$ 
depending on the order of
the zeros. For the  simple zeros 
the value of the discontinuity through any continuous path 
of gauge fields crossing the
nodal field is   $\pi $. For
double zeros the discontinuity  is $2\pi$ and so forth. 
 If the trajectories of fields
correspond to  paths of three dimensional gauge fields
induced by  four-dimensional gauge fields with
non-trivial topological charge $q$ it can be shown from the index theorem 
that  the total discontinuity  along the trajectory will be 
equal to $2\pi q$ \wwitten.  

 The regularized value of the fermionic determinant
in the transition regime has an imaginary component of the effective actions
which undergoes non-integer discontinuities. This fact 
signals an extra degree of 
non-analyticity ({no holomorphy}) of the determinant in the transition
regime and, thus, indicates
 that there is a radical difference of the transition regime with the other
regimes. 
It  has a completely different new physical behaviour which 
in any case cannot be interpreted in pure analytic terms.

Moreover,
because of the parity symmetry of  Dirac operator,  if  A is a nodal
gauge field its transformation under parity $A^P$ is also a nodal configuration which
implies that the singularities are invariant under parity transformation. 
Thus, the whole non-analytic  component  $ h(A)$   is parity preserving, i.e. 
\eqn\ph{
h\lbrack A^P\rbrack = h\lbrack A\rbrack.
}
The true source of parity symmetry breaking has a perturbative origin:
 the induced Chern-Simons term.
Although the Chern-Simons radiative corrections can removed by a local counterterm, 
 gauge invariance under large gauge transformation is broken if $k_R\notin \IR$.
Therefore the theory cannot be parity invariant and gauge invariant at the same
time in this case. Only in the case $k_R=2\pi n$ both symmetries can be
simultaneously preserved.

In other terms, because of the hermiticity of $\Dsla$ all eigenvalues are
real. Thus, the only source of imaginary terms in the effective action comes
from negative eigenvalues $-1= {\rm e}^{i\pi}$ . Now at nodal points one positive
eigenvalue becomes negative or one negative eigenvalue becomes positive. 
Thus, generically,  $ h(A)$ has a $\pi$ discontinuity at configurations with
fermionic zero modes. In the case of fermions in the adjoint representation,
there is a level crossing at nodal points between eigenvalues becoming positive and
eigenvalues becoming negative (see Figure 3). This explains why in that case the
singularities of the effective action do not have any physical effect.

\begin{figure}
\centerline{\epsfxsize=12cm \epsfbox{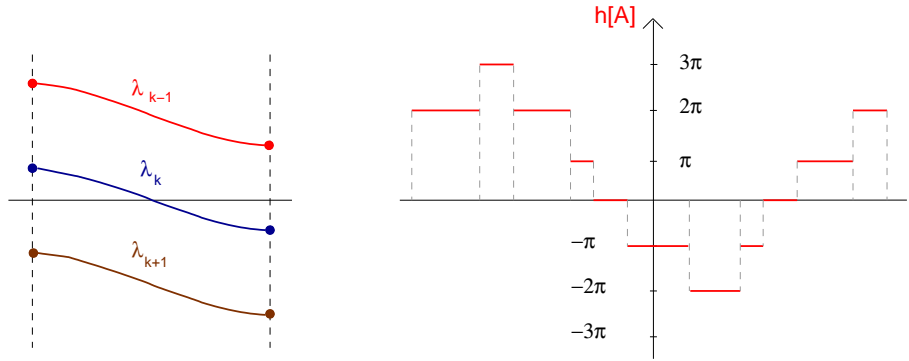}}

\caption{{\bf } Spectral flow and singular behaviour of the effective
action of a fundamental fermionic determinant}
\end{figure}

In all  gauge invariant regularizations  the non-analytic term of the effective
action $h(A)$  is proportional to $k_R-k$ as a  consequence of gauge 
symmetry. This means that the effective 
counting of zero-level crossings becomes regularization dependent.

In order to illustrate the phenomenon let us consider a lower dimensional
example: a fermionic quantum rotor under the action of a magnetic flux
\dunne\jllp. In one dimension the equivalent of Chern-Simons action is 
$${   k}\ S_{\rm cs}={k}\int A= {   k}\, \epsilon $$
and the fermionic determinant of $\Dsla=d_\theta+A_\theta$
 can be exactly computed \jllp\
$$\det \Dsla = {\rm e}^{- \Gamma(A)}$$
with
\eqn\rot{
\eqalign{
\Gamma(A)&= -\log\Bigl(
|\cos {\epsilon\over 2}|+ i\, 2\pi\, {   k_R}\,\left[{\epsilon\over 2\pi}+ {1\over 2}\right]
-i\pi k_R\Bigr) \cr
&\cr
&=-\log\Bigl\lbrack |\cos {\epsilon\over 2} |+ i \, 2\pi\, {   k_R }\, \left({\epsilon\over 2\pi} - 
{\rm Int}
\left({\epsilon\over 2\pi}+ {1\over 2}\right)\right)
\Bigr\rbrack}
}
with the renormalized coupling constant
$$k_{\rm R}= \begin{cases}
{1\over 2}& {{\rm if}\  m > 2n+1/2}\cr
 {    {1 \over\pi}
\arctan{\lambda_+\over\lambda_- }}& {{\rm if}\     m = 2n+1/2}\cr
{0} &{{\rm if}\   m < 2n+1/2 }\end{cases}$$
depending on regularization parameters $m,n$ and $\lambda_+,\lambda_-$.   
${\rm Int}(x)$ denotes the integer part of $x$.
The effective action \rot, is gauge invariant for any value of these parameters.
The Chern-Simons term  of the imaginary of the effective action
$  {   k_R}\ S_{\rm cs}=k_R \, {\epsilon}$ is compensated by a non-analytic component
\eqn\rott{
h(A)= { -  k_R\pi  }\, \left[ 
2 \,{\rm Int}
\left({\epsilon\over 2\pi}+ {1\over 2}\right)\right]}
which is parity invariant but transforms under global gauge transformations
in a way that compensates the anomalous transformation of the Chern-Simons
part. Notice that the whole imaginary part of $\Gamma(A)$ is  proportional to $k_R$ 
in all regularization regimes.
\medskip
%

In fermionic determinants the interpretation of singularities in terms of
nodal configuration is quite natural. However, in  Chern-Simons theory
the divergence of $\Gamma_A$ at  one-loop order  is more intriguing because there
is not an apriori reason for  singularities. In the
Schr\"odinger representation physical states  are described by
functionals of gauge fields which as pointed out in \kar\afll\ in Chern-Simons theory 
vanish at certain nodal configurations.
It is therefore not unreasonable that the effective action of the theory 
could diverge at some classical configurations which might be related to
nodes of the  vacuum state. In general, this type of singularity indicates a 
suppression of  tunneling. One can identify some configurations where the
one loop effective action diverges. In fact, it is easy to show that
 there is a discontinuity of $h(A)$ at the sphaleron gauge field on $S^3$.
This is a gauge field which is a saddle point of Yang-Mills action, i.e.
it satisfies the Euclidean Yang-Mills equations and Bianchi identity
\eqn\bym{D_A F(A)=D_A^\ast F(A)=0.}
 It is given explicitly   by
$$
\left[A_{{\rm{sph}}}\right]_j={ 4R\over (x^2+4 R^2)^2}\left(
4R \epsilon^a
_{jk} x^k - 2x^a x_j+ [x^2-4 R^2]\delta^a_j\right)\sigma_a
$$
for  $SU(2)$ gauge fields ($R$ is the radius of the $S^ 3$ sphere).
The proof that $A_{{\rm{sph}}}$ is a nodal point follows from  equations \bym
which imply  that  $\ast F(A)$ is a zero mode of the operator $ \Delta_A$  which  generates the 
one-loop corrections of the \cst\ \ddel. 
There exist a similar phenomenon for massless fermions on the adjoint
representation. The sphaleron is a nodal configuration of the corresponding
determinant with the same spectral flow. Now, since the fermions are in the 
adjoint representation, two level cross the zero level at the sphaleron 
configuration (see Fig. 3).

\begin{figure}
\centerline{\epsfxsize=7cm \epsfbox{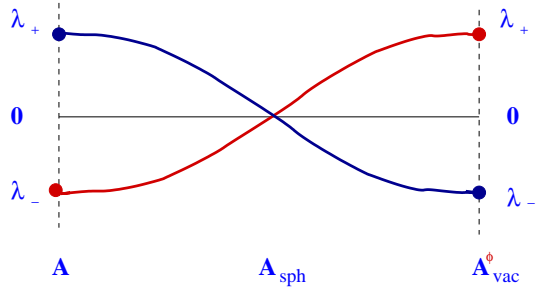}}

\caption{{\bf } Spectral flow  of the
fermionic determinant in the adjoint representation}\end{figure}

In this case  the fermionic determinant 
$\det \Dsla$ has the same properties that the vacuum
state of $3+1$ dimensional gauge theories at    $\theta=\pi$
and the discontinuity of $h(A)$ at sphaleron
configurations on $S^3$ is a physical property which encodes the 
tunneling suppression due to the effect of massless fermions \affs.

The dependence of the fermionic determinant on the background
gauge field contributes to the understanding of the role of singular
contributions in the effective action. The existence of
 zero modes determine the existence of discontinuities
in the imaginary part of the effective action. In the real part the
singularities are more severe. At nodal configurations the real part of the
effective action becomes infinite signaling the failure of perturbation theory
and the vanishing of the corresponding determinant.  
The analysis of the physical role of these singularities in Chern-Simons theory
and its possible survival at higher orders in the coupling constant $1/k$ 
is an open problem.

 In order to obtain a better physical picture of the transition regime let us
analyse the case of  pure Abelian Chern-Simons theory.
\eqn\aba{
S_{\rm cs}={   k_R\over 4 \pi}\int A\land F(A)}
Although in general there is no quantization condition for the 
coupling constant $k_R$, in the presence of  magnetic monopoles
in  $M= S^1\times {\af T}^2$, consistency requires   its quantization.

In temporal gauge and flat gauge fields the effective action \aba\ reduces to
$$
S_{\rm cs}={   k_R \pi}\epsilon^{ij} \int a_i\dot{a}_j
$$
and can be quantized as a quantum Hall effect in a dual torus $\widehat{\af T}^2$
with magnetic charge $k_R\in\IZ$.
The number of physical states is finite and equals  the value of the magnetic charge
$k_R$\bosnair. This explains why $k_R$ should be quantized.

In  the transition   regime  a massless
fermion induces an effective action with $k_R\notin \Z$ and extra
non-analytic terms in the imaginary part.  In order to analyze the physical effect of
these terms let us consider a slightly different action with a similar basic behaviour
$$S_{\rm cs}={   k_R \pi}\epsilon^{ij} \int [a_i]\, \dot{a}_j
$$
where $[x]=x-{ }{\rm Int}({x})$  and $k_R\notin \IZ$. The
 system governed by such an action is equivalent to a charged particle moving  in a torus 
under the action of two magnetic fields: one uniform magnetic field with non-integer
total magnetic flux $k_R$ across the torus , and an extra magnetic field
with a  delta-like singularity  whose magnetic flux
just cancels that of the uniform magnetic field. 
$$
F_{12}={k_R\pi}\left[2-\delta\left({a_1}\right)-\delta\left({a_2}\right)\right].
$$
Thus, the 
total magnetic flux is null and 
gauge invariance under large gauge transformations is restored.

The quantum system has in this case only one vacuum state. Thus,
the physical regime associated to transition regularization may  be very
different from the one obtained from generic regularization schemes.
This  would explain in physical terms the smooth interpolation 
 between the two generic  regularization regimes through  
the transition regime. 

The fact that different regularizations of the theory give
rise to different quantum theories is not so surprising. One simple but 
paradigmatic example is topological quantum
mechanics
on a Riemann surface $\Sigma$ of genus $h$ in the
presence of a magnetic field $A$ with
magnetic charge $k$ \kar. In standard Hamiltonian 
formalism
 the quantum Hamiltonian is trivial ($H=0$)  as
corresponds to a topological theory and
the dimension of the space of quantum
states is finite and given by 
$
\dim {{\mathcal H}}_k^0= 1-h + k,$
 for  $k> h-1$.
 However, if
the theory is   regularized  by means of a metric dependent kinetic term,
$$L(x,\dot x)= {1\over 2 \Lambda} g_{ij}\dot{x}^i\dot{x}^j+
 A_i \dot{x}^i,$$
the Hamiltonian becomes
$H_{\Lambda}={\Lambda\over 2}\Delta_A^g$, and the topological limit
$\Lambda\to \infty$ is governed by
the ground states of $H_{\Lambda}$.
The quantum Hilbert space of the  topological field theory
obtained by this method can have a dimension lower than $1-h+k$, depending on
the symmetries of the background metric
$g$  of  the regularization  \kar. In particular, this is the case when the metric $g$  breaks the
degeneracy of the ground state of the covariant Laplacian
$\Delta_A^g$. The standard result is obtained by choosing 
 only metrics  which is compatible with the magnetic
field $B=dA$, in the sense that they give rise to 
a K\"ahler structure on $\Sigma$.

\newsec{A c--theorem in three-dimensions }

The existence of  different regimes in the  ultraviolet regularization in  
Chern-Simons theory also has relevant implications for the induced gravitational interactions.
Although Chern-Simons action is metric independent  the quantum
corrections generate a  finite gravitational \cs\ term
\eqn\csg{S_{\rm csg}= {\kappa\over 4\pi}\int \left[\epsilon^{\mu\nu\sigma} R_{\mu\nu a b}\omega^{ab}_\sigma+{2\over 3}
\omega^{b}_{\mu a}\omega^{c}_{\nu a}\omega^{a}_{\sigma c}\right].}
 This term which gives rise to a  metric 
independent effective action can  be canceled by the introduction of a local 
counterterm. But then a  framing anomaly is generated
as a physical effect of the theory \usb. The novel effect is that this anomaly
 also become dependent  on the regularization regime as the parity anomaly.

 The  induced gravitational  \cs\ term  was conjectured to be of the 
form $\kappa= {c/24}$, where $c$ is the central charge  of the  conformal
theory associated to Chern-Simons theory \wittten. In the present
 case, $ c=k(N^2-1)
/(k+N)$. 
In perturbation theory, this means that 
$$\kappa={N^2-1\over 24} \sum_{n=0}^{\infty} (-{N\over k})^n .$$
However, as anticipated $\kappa$  depends on the choice of regularization regime. 
The  one loop contribution
$$\kappa^{[1]}_{\rm R}= \begin{cases}
{N^2-1\over 24}& {{\rm if}\  m > 2n+1/2}\cr
 {    {N^2-1 \over 12\pi}
\arctan{\lambda_+\over\lambda_- }} & {{\rm if}\     m = 2n+1/2}\cr
{0} &{{\rm if}\   m < 2n+1/2 }\end{cases}$$
agrees with the expected value $\kappa=(N^2-1)/24 $ only if $m> 2n+ \ha$.
  The vanishing of  $\kappa$ in the  regime with $m< 2n+ \ha$   was first
anticipated  by  Witten \wittten. In this scheme a second order perturbative
calculation was carried out  in  Refs. \AS, and the result
seems to agree with the standard case. In the transition regime $\kappa$ 
 depends on the weights $\lambda_+$ and $\lambda_-$ of the parity odd and parity even
regulators and does not correspond to any previously expected
behaviour\foot{If  the scalar laplacian were considered 
instead of the vector laplacian the result for the transition regime will be
different}. In this case there is relation between the value of $\kappa$
and the renormalized \cs\ coupling constant $k_R$ 
\usb,
$$\kappa ={(k_R-k)(N^2-1)\over 24 N}.$$ 
The above results suggest that this relation holds for the three regimes. 
It would be very interesting
to investigate if the property also holds beyond one loop approximation.

Other types of gravitational terms like Einstein or cosmological
constant terms  can also be generated  in the effective action, but
  they present  linear or cubic UV divergences which need to 
be renormalized leaving an extra ambiguity in the actual values
of the corresponding renormalized couplings. Metric independence 
requires the cancellation of  both couplings. But 
the same gravitational \cs\ term also contains 
some hidden Einstein and cosmological terms  
when the gauge field is written in terms
of the {\it vierbein} and the spin connexion \achucarro. 
The
different values of the renormalized gravitational \cs\ constant
also adds an extra source of metric dependence.
Although the induced Chern-Simons term can be removed by a choice
of renormalization scheme its non-analytic counterpart cannot  and in fact
yields an extra frame dependent contribution.
 Only the third regime provides a fully consistent metric independent theory
without parity and framing anomalies.

This connection between the renormalization of Chern-Simons coupling and
the induced gravitational Chern-Simons coefficient ant its  relation to the 
central charge of the associated conformal theory
suggests   a possible  extension of  Zamolodchikov's c-theorem to three-dimensional
systems. Topological Chern-Simons theories would correspond to two
dimensional conformal theories and the interpolating regularized topologically 
massive theories will generate a flow from one theory with one Chern-Simons 
coupling to another with a different one. A c-theorem would establish the
existence  of a monotone function along this renormalization group flow which
will coincide with the coupling of  gravitational \cs\ term at  topological fixed points. One
natural candidate for  Zamolodchivov c-function, thus, can be defined in terms of  the
induced gravitational Chern-Simons  term which is identical to $\kappa$ at the
pure Chern-Simons theories and   varies  along renormalization 
group trajectories. A concrete proposal based on an version of Zamolodchikov
theorem formulated in Ref. \flc\ can be established  from the  following
spectral representation of the stress tensor  correlators
\eqn\cth{\eqalign{
 \phantom{asdfadfadf}&{\Bigl\langle\!\!\!\Bigl\langle} T_{\alpha\beta}(x)  T_{\mu\nu}(0)
{\Bigl\rangle\!\!\!\Bigr\rangle_{\rm odd}} =\cr 
&\cr
&\displaystyle 
- {1\over 192 \pi}\int d^3 x {{\rm e}^{i p.x}\over (2\pi)^{3/2}} 
\int_0^\infty d\lambda \ 
{ \lambda \, {  {c(\lambda)}}
\over \sqrt{ p^2+\lambda^2}}\cr
&
\cr
\displaystyle
&\left[{\epsilon_{\mu\sigma\alpha}}\,p^\sigma 
(p_{\nu}p_{\beta}- \delta_{\nu\beta})+ \right.
 { \epsilon_{\nu\sigma\alpha}}\,p^\sigma (p_{\mu}p_{\beta}- \delta_{\mu\beta})
+ 
\cr
&
\cr
&\left.
\displaystyle
  {\epsilon_{\nu\sigma\beta}}\,p^\sigma (p_{\mu}p_{\alpha}- \delta_{\mu\alpha})+
 {\epsilon_{\mu\sigma\beta}}\,p^\sigma (p_{\nu}p_{\alpha}- \delta_{\nu\alpha})\right]
}}
   ${ c(\lambda)}$ emerges as a natural candidate for a Zamolodchikov c-function for
three-dimensional theories. Unfortunately,  $c(\lambda)$ cannot be universally 
monotone for the same reasons  as the similar spectral representation of the
flow  of the effective $k$ coupling cannot be monotone in all regularization 
regimes of pure Chern-Simons theory (see Fig.1)  \foot{This observation was 
made by Ian Kogan and one of us (M.A.)}. This negative
result does not exclude the existence of another extension of c-theorem to 2+1
dimensional  theories. It merely points out that the spectral representation of
the gravitational Chern-Simons term is not a good c-function.  On the other
hand for purely bosonic theories there are not  axial Chern-Simons  like
interactions which could generate a simple way of describing the irreversibility
of the renormalization group flow. 

\newsec{ Discussion}

The presence of non-analytic terms in the effective action is
fundamental for the right physical description of \cst\  and massless fermions in 2+1 dimensions.
The existence of such  contributions is pointed out by the appearance of 
different perturbative corrections in different gauge invariant
regularizations.
The discontinuities associated to those non-analytic terms signal the presence
of physical singularities associated to the zeros of the partition function in
some backgrounds. The appearance of nodal configurations is also related to
the quantum tunneling suppression. In massless fermions nodal contributions
are associated to the existence of fermionic zero modes. However, the structure
of  singularities and discontinuities depends on the regularization regime and
makes possible the physical differences between the corresponding
 theories. In regularizations with
 transition regimes the nature of the singularities associated to non-analytic
terms suggest a non holomorphic behaviour of the effective partition function
in terms of classical fields. It is remarkable that it is possible
to find out  gauge invariant regularization regimes where there are not parity and framing
anomalies. This suggests that those anomalies can be better understood as
spontaneous symmetry breaking phenomena rather than genuine anomalies.
In some toy models it has been shown that the transition regularizations keep constant
the number of physical states.
It would be very interesting to analyze the behaviour of the number of 
physical states of the \cst\ in the transition regime and verify if  it is
dramatically reduced as in the toy model. Finally, it is pointed out why
the extension of Zamolodchikov c-theorem to three-dimensional 
theories is  an interesting still open problem.


\newsec{ Acknowledgments}
We thank Fernando Falceto, Gloria Luz\'on and very specially Ian Kogan for enlightening discussions during
different stages of the realization of the paper. This work is partially
supported by  CICYT (grant 
FPA2004-02948).

\end{document}